\documentclass[sigconf]{acmart}

\usepackage{amscd,amsfonts,amsbsy,rotating}
\usepackage{balance}
\usepackage{graphicx}
\usepackage{epsfig,epstopdf}
\usepackage{subfigure}
\usepackage{multirow}
\usepackage{booktabs}
\usepackage{color,xcolor}
\usepackage{url}
\usepackage{latexsym,bm}
\usepackage{enumitem,balance,mathtools}
\usepackage{wrapfig}
\usepackage{euscript}
\usepackage{algorithm}
\usepackage{algorithmic}
\usepackage{ifpdf}
\usepackage{diagbox}
\usepackage{caption}
\usepackage{makecell}
\usepackage{subfigure}
\usepackage{bm}
\usepackage{appendix}

\newcommand{\bs}{\boldsymbol}

\newcommand{\bx}{\bs{x}}

\newcommand{\br}{\bs{r}}
\newcommand{\btheta}{\bs{\theta}}

\newcommand{\be}{\bs{e}}
\newcommand{\bv}{\bs{v}}

\usepackage{array}
\newcolumntype{N}{@{}m{0pt}@{}}

\newcommand{\minisection}[1]{\vspace{5pt}\noindent\textbf{#1.}}

\AtBeginDocument{%
  \providecommand\BibTeX{{%
    \normalfont B\kern-0.5em{\scshape i\kern-0.25em b}\kern-0.8em\TeX}}}

\acmSubmissionID{rt0873o}

\begin{document}

\copyrightyear{2019} 
\acmYear{2019} 
\setcopyright{acmcopyright}
\acmConference[KDD '19]{The 25th ACM SIGKDD Conference on Knowledge Discovery and Data Mining}{August 4--8, 2019}{Anchorage, AK, USA}
\acmBooktitle{The 25th ACM SIGKDD Conference on Knowledge Discovery and Data Mining (KDD '19), August 4--8, 2019, Anchorage, AK, USA}
\acmPrice{15.00}
\acmDOI{10.1145/3292500.3330870}
\acmISBN{978-1-4503-6201-6/19/08}

\settopmatter{printacmref=true}

\title{Deep Landscape Forecasting for Real-time Bidding Advertising}

\author{Kan Ren, Jiarui Qin, Lei Zheng, Zhengyu Yang, Weinan Zhang, Yong Yu}
\affiliation{
  \institution{Shanghai Jiao Tong University\\
    \{kren, qinjr, zhenglei, zyyang, wnzhang, yyu\}@apex.sjtu.edu.cn\\~}
}

\renewcommand{\shortauthors}{K. Ren, et al.}

\begin{abstract}
The emergence of real-time auction in online advertising has drawn huge attention of modeling the market competition, i.e., \textit{bid landscape forecasting}. The problem is formulated as to forecast the probability distribution of market price for each ad auction. With the consideration of the censorship issue which is caused by the second-price auction mechanism, many researchers have devoted their efforts on bid landscape forecasting by incorporating survival analysis from medical research field. However, most existing solutions mainly focus on either counting-based statistics of the segmented sample clusters, or learning a parameterized model based on some heuristic assumptions of distribution forms. Moreover, they neither consider the sequential patterns of the feature over the price space. In order to capture more sophisticated yet flexible patterns at fine-grained level of the data, we propose a Deep Landscape Forecasting (DLF) model which combines deep learning for probability distribution forecasting and survival analysis for censorship handling. Specifically, we utilize a recurrent neural network to flexibly model the \textit{conditional} winning probability w.r.t. each bid price. Then we conduct the bid landscape forecasting through probability chain rule with strict mathematical derivations. And, in an end-to-end manner, we optimize the model by minimizing two negative likelihood losses with comprehensive motivations. Without any specific assumption for the distribution form of bid landscape, our model shows great advantages over previous works on fitting various sophisticated market price distributions. In the experiments over two large-scale real-world datasets, our model significantly outperforms the state-of-the-art solutions under various metrics.
\end{abstract}



\maketitle

\section{Introduction}\label{sec:intro}
Emerged from 2009 \cite{muthukrishnan2009ad}, real-time bidding (RTB) has become one of the most important media buying mechanisms in online advertising.
In RTB, the advertisers propose the bid price in real time, according to their own bidding strategies and the auction side information \cite{zhang2014optimal}, then the ad exchange decides the winner in the market of the auction, i.e., the advertiser with the highest bidding price in this auction \cite{yuan2013real}.
From the view of advertiser, specifically, the bid price is decided according to the estimated \textit{utility} \cite{ren2016user} and the \textit{cost} of the given auction request \cite{ren2018bid,ren2016user,wu2015predicting}.
On one hand, the utility generally represents the positive user response probability such as click-through rate (CTR) or conversion rate (CVR).
On the other hand, the cost is the price which the advertiser would probably pay for the given auction.
  
\begin{figure}[t]
  \centering
  \includegraphics[width=0.8\columnwidth]{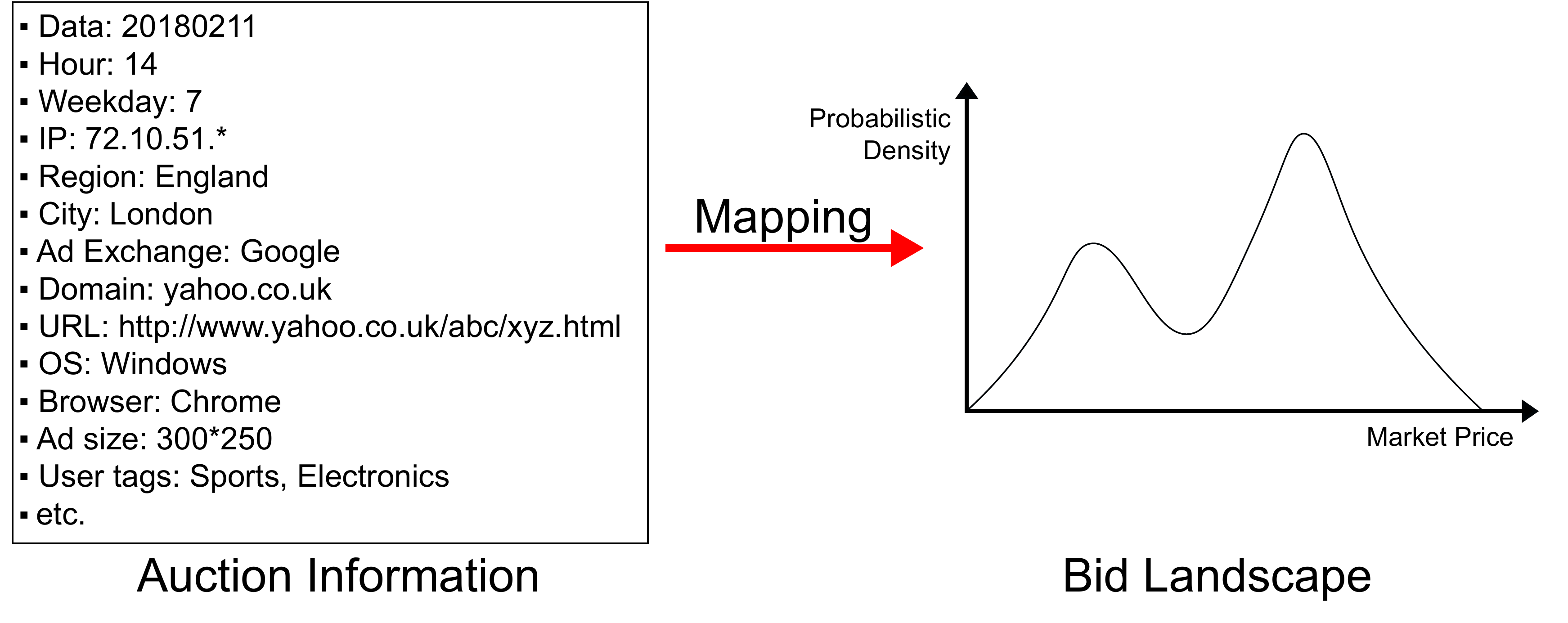}
  \caption{The main task of the bid landscape forecasting.}\label{fig:task}
  \vspace{-12pt}
\end{figure}

Note that, the true charge for the winner of the auction is the highest bid price from her competitors, which is defined as the \textit{market price}\footnote{The terms `market price' and `winning (bid) price' are used interchangeably in the related literatures \cite{amin2012budget},\cite{cui2011bid},\cite{wu2015predicting}. In this paper, we use `market price'.} in the second-price auction mechanism.
So that, from an advertiser's perspective, predicting the market price is a crucial but challenging problem since the highest bid from hundreds or even thousands of advertisers for a specific ad impression is highly dynamic and almost impossible to predict by modeling each advertiser's strategy \cite{amin2012budget}. 
Moreover, only with the market price distribution, the advertiser can estimate the corresponding \textit{winning probability} given an arbitrary bid price, which supports the subsequent bid decision making \cite{zhang2014optimal,ren2018bid}. 
For example, \citet{lin2016combining} adopted a bidding strategy by proposing the bid price according to the estimated market price and \citet{ren2018bid} presented a method of learning bidding strategy through optimizing advertisers' profits which requires the forecasted bid landscape for each bid request sample.
Thus, it is more practical to model the market price as a stochastic variable \cite{wu2015predicting,wu2018deep} and predict its distribution given each ad request feature, named as \textit{bid landscape forecasting} \cite{wang2016functional,cui2011bid} and illustrated in Figure~\ref{fig:task}.

The previous works on bid landscape forecasting can be divided as two streams.
The first stream is mainly based on statistically counting from the segmented samples, for example counting per campaign \cite{zhang2016bidaware} or by some particular attribute combinations \cite{wang2016functional}. Different samples in the same segment share the same market price distribution which is too coarse-grained and often result in low prediction performance.
The second stream is based on predefining a parameterized distribution form, such as log-normal distribution \cite{cui2011bid}, Gaussian distribution \cite{wu2015predicting,wu2018deep} or Gamma distribution \cite{zhu2017gamma}, and then learning the distribution parameters with the observed data.
However, as is discussed in \cite{yuan2014empirical} these assumptions are often too restricted and rejected by statistical tests thus lack of generalization.

Yet there is another challenge of the bid landscape foresting which is the \textit{censorship} issue.
Since RTB adopts the one-slot second-price auction mechanism \cite{yuan2013real}, 
only the winner, who submits the highest bid price, will know the market price, i.e., the charged price, while others can only know that the market price is higher than their bids, which is called right-censored data \cite{wu2015predicting,wang2016functional}.
To handle this censorship, many works \cite{zhang2016bidaware,wang2016functional,wu2015predicting,wu2018deep} borrow the idea of survival analysis in medical data science and take the losing logs into consideration to better model the true market price distribution.
However, these methods rely on only the losing logs and do not take a comprehensive view of considering both winning logs and losing logs for censorship handling.

In this paper, we propose a deep neural network methodology name as Deep Landscape Forecasting (DLF) model, without any presumed heuristic forms of market price distribution, to better capture the sophisticated patterns for each auction in bid landscape forecasting.
Specifically, we utilize a recurrent neural network to model the conditional probability of the winning event given a bid.
And then the model forecasts the distribution of market price by probability chain rule and naturally derive the winning probability distribution of arbitrary bid prices, for the given auction.
We not only train the model through maximizing the log-likelihood of the true market price in the winning logs.
Moreover, we also adopt a comprehensive loss function over both winning logs and losing logs, to handle the censorship.

The novelty of our methodology are threefold.
\begin{itemize}[leftmargin=3mm]
  \item \textbf{Fine-grained prediction}: Rather than forecasting the bid landscape over sample segments \cite{zhang2016bidaware,wang2016functional}, our method can predict the ``personalized'' market price distribution and the corresponding winning probability distribution for each specific bid request.
  \item \textbf{No assumption of distribution forms}: Based on the novel modeling methodology, our model manages to generate flexible forecasting results for each ad request without making any prior assumptions about the market price distribution, which will be illustrated in the experiment.
  \item \textbf{Novel Censorship loss function}: We adopt a comprehensive loss function for censorship handling and make a step further upon the traditional survival analysis methodologies \cite{wu2015predicting} to better model the market price distribution.
\end{itemize}
To our knowledge, this is the first work that proposes an end-to-end deep learning model without any distributional assumptions for bid landscape forecasting.
To deal with the right-censored problem, we use both the observed winning data and censored losing data to derive an unbiased learning. 
In addition, the experimental results over two real-world datasets show the significant improvement of our model over strong baselines under various evaluation metrics.

\section{Related Works}\label{sec:related-work}

\minisection{Bid Landscape Forecasting} As is discussed in the above section, bid landscape forecasting has become an important component in RTB advertising and drawn much attention in recent works \cite{ren2016user,ren2018bid,lin2016combining}.

\begin{figure}[t]
  \centering
  \includegraphics[width=1.0\columnwidth]{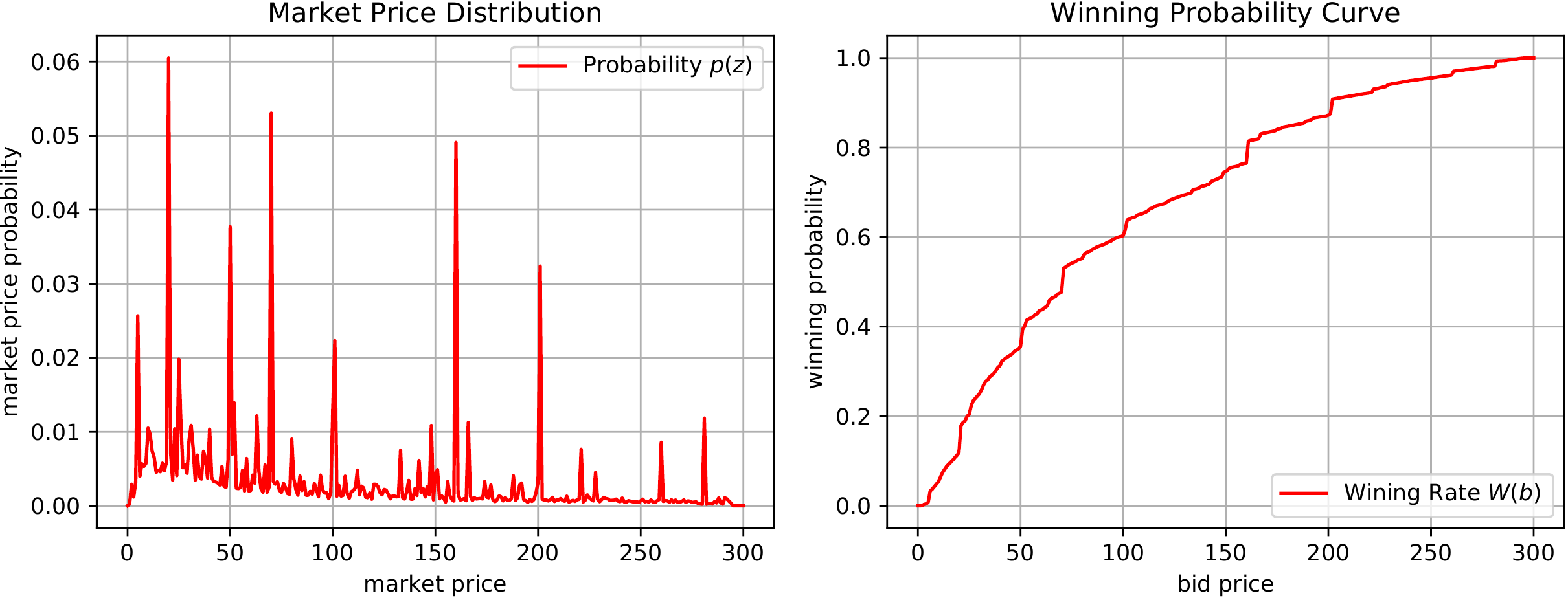}
  \caption{Market price distribution and winning function.}\label{fig:mp_distribution}
  \vspace{-10pt}
\end{figure}

In the view of distribution modeling methods, there are two phases.
In the early phase, researchers proposed several heuristic forms of functions to model the market price distribution. In \cite{zhang2014optimal,ren2018bid,ren2016user}, the authors provided some analytic forms of winning probability w.r.t. the bid price applied on the campaign level, which is based on the observation of the winning logs.
Later in the recent researches, some well-studied distributions are applied in market price modeling.
\citet{cui2011bid} presented a log-normal distribution to model the market price ground truth.
\citet{wu2015predicting} proposed a regression model based on Gaussian distribution to fit the market price.
Recently, Gamma distribution for market price modeling has also been studied in the work \cite{zhu2017gamma}.
The main drawback of these distributional methods is that these restricted empirical preassumptions may lose the effectiveness of handling various dynamic data and they even ignore the sophisticated real data divergence as we show in Figure~\ref{fig:mp_distribution}.

In the view of forecasting, the goal is to predict the true market price distribution, i.e., bid landscape, for the given auction sample.
A template-based method was present in \cite{cui2011bid} to fetch the corresponding market price distribution w.r.t. the given auction request.
\citet{wang2016functional} proposed a clustering-based tree model to automatically segment the data samples and built a non-parametric estimator for each leaf node to predict the market price distribution.
These methods can only perform coarse-grained bid landscape forecasting based on each data segment which misses the individual patterns of each ad request.
The authors in \cite{wu2015predicting} proposed a linear regression method to model the market price w.r.t. auction features.
Nevertheless, those methods do not care much about the non-linear patterns in the real data, i.e., such as the co-occurrence and correlations between features \cite{qu2016product} and the similarity/distinction among data segments, which may result in poor forecasting performance on different ad campaigns.
Recently, \citet{wu2018deep} have proposed one work about winning price prediction with deep models.
However, they use several assumptions about the form of market price distribution which is not flexible in practice.
Moreover, as is stated in their paper, the goal of their method is to directly predict the winning price, rather than forecasting the bid landscape, thus their method does not outperform the tree-based model \cite{wang2016functional} over the log-likelihood metric which is one of the main evaluation metrics.

In this paper, we focus on fine-grained bid landscape forecasting at impression-level, and utilize deep recurrent neural network to flexibly model the non-linear and sequential patterns in market price distribution without any prior assumption of the distribution form.

\minisection{Learning over Censored Data}
The data censorship is another challenge for bid landscape forecasting.
In the online advertising field, many models based on survival analysis have been studied so far.
\citet{wu2015predicting,wu2018deep} proposed a censored regression model using the lost auction data to alleviate the data bias problem.
Nevertheless, the Gaussian distribution or other distributional assumptions \cite{zhu2017gamma} turn out to be too restricted while lacking of flexibility for modeling sophisticated yet practical distributions.
Another problem is that these regression models \cite{wu2015predicting,wu2018deep,zhu2017gamma} can only provide a point estimation, i.e., the expectation of the market price without standard deviation, which fails to provide winning probability estimation given an arbitrary bid price to support the subsequent bidding decision \cite{ren2018bid}.
\citet{amin2012budget} implemented Kaplan-Meier estimator \cite{kaplan1958nonparametric} for handling the data censorship in sponsored search.
Kaplan-Meier estimator is a classic method in survival analysis which deals the right censored data in medical research \cite{gordon1985tree,ranganath2016deep}.
The authors of \cite{wang2016functional,zhang2016bidaware} also utilized this non-parametric estimator to predict the winning probability.
However, Kaplan-Meier estimator is merely statistically counting on the segmented data samples, thus fails to provide a fine-grained estimation, i.e., prediction on a single ad auction level.

Another school of survival analysis methods is Cox proportional hazard model \cite{cox1992regression}.
This method commonly assumes that the instant hazard rate of event (i.e., auction winning in our case) occurrence is based on a base distribution multiplied by an exponential tuning factor.
Recent works including \cite{katzman2016deep,luck2017deep,park2007l1} all used the Cox model with predefined base function to model the hazard rate of each sample, such as Weibull distribution, log-normal distribution or log-logistic distribution \cite{lee2003statistical}.
However, the problem is that the strong assumption of the data distribution may result in poor generalization in real-world data.

Considering all the limitations above, we propose our deep landscape forecasting method which models the conditional winning probability given the sequential price patterns through recurrent neural network, and maximizes the partial likelihood over the historic winning and losing data.
Note that our model takes end-to-end learning in a unified learning objective without making any distributional assumptions, which is capable of fitting various bid landscape data and providing fine-grained prediction for each ad impression.

\minisection{Recurrent Neural Network} 
Due to its adequate model capability and the support of big data, deep learning, a.k.a. deep neural network, has drawn great attention.
Among them, recurrent neural network (RNN) whose idea was firstly developed two decades ago shows satisfying performance for sequential pattern mining.
And its variants like long short-term memory (LSTM) \cite{hochreiter1997long} employ memory structures to better captures dynamic sequential patterns.
In this paper we borrow the idea of RNN for modeling conditional winning probability.
And in the experiments, our model shows superior better performance against state-of-the-art baselines including survival analysis methods \cite{katzman2016deep,lee2018deephit} and recent bid landscape forecasting models \cite{wang2016functional,wu2018deep}.

\section{Methodology}\label{sec:method}
In this section, we firstly present the preliminaries used in our discussion in Section~\ref{sec:preliminaries} and formulate the problem in Section~\ref{sec:prob-def}.
Then in Section~\ref{sec:DLF} we discuss our bid landscape forecasting model in detail.
We conduct a deep analysis of the model in Section~\ref{sec:model-realization}.

\subsection{Preliminaries}\label{sec:preliminaries}


In RTB scenario, the advertiser is asked to propose a bid price $b$ after receiving a bid request $\bx$ for auction.
The bid request contains three parts of the auction information including user (e.g., location, browser label, etc.), publisher (e.g., web URL and ad slot size, etc.) and the ad content (e.g., product type, time and creative content).
The goal of the advertiser is to propose an appropriate bid price and win the auction in a cost-effective manner \cite{ren2018bid,yuan2013real}.

One of the challenges is that it is infeasible to model the bidding strategy of each competing bidder since the participating advertisers do not interact with each other \cite{wang2016functional}.
It is natural to model the market as a whole and regard the market price as a variable $z$ \cite{wang2016functional,lin2016combining,zhang2014optimal}.
Recall that the market price is the second highest bid price among all the bidders in the second-price auction, i.e., the highest bid price from the competitors in the view of the auction winner.
The probability density function (P.D.F.) of the market price $z$ is $p(z), z>0$.

Now that we have the P.D.F. $p(z)$ of the market price $z$, we can derive the winning probability of proposing the price at $b$ as
\begin{equation}
W(b) \doteq \text{Pr}(z < b) = \int_0^b p(z) dz ~, \label{eq:win-prob-definition}
\end{equation}
which is the probability that our bid price is larger than the market price, i.e., winning the auction.
Then the straightforward definition of the ``losing'' function is
\begin{equation}
S(b) \doteq \text{Pr}(z \geq b) = 1 - W(b) = \int_b^{\infty} p(z) dz ~, \label{eq:survival-definition}
\end{equation}
which represents the losing probability of proposing the bid price $b$.
Note that in survival analysis \cite{katzman2016deep,Li2016mtlsa}, the market price is regarded as the patient's underlying survival period and the bid price is the investigation period, thus \textbf{winning} and \textbf{losing} the ad auction respond to the ``death'' and ``survival'' status of one patient \cite{zhang2016bidaware}.

\subsection{Problem Definition}\label{sec:prob-def}
The data of the bidding logs are represented as a set of triples $\{(\bx, b, z)\}$, where $\bx$ is the feature of the bid request, $b$ is the proposed bid price in that auction.
Here $z$ is the observed market price if the advertiser previously won this auction and she has already known the true market price, but $z$ is unknown (and we marked $z$ as null) for those losing auctions.

The main problem of bid landscape forecasting is to estimate the probability distribution $p(z | \bx)$ of the market price $z$ with regard to the bid request feature $\bx$.
Formally speaking, the derived model is a ``mapping'' function which learns the patterns within the data and predicts the market price distribution of each auction as
\begin{equation}\label{eq:task-function}
p(z | \bx) = T(\bx) ~.
\end{equation}
The general goal has been illustrated in Figure~\ref{fig:task}.

\subsection{Deep Landscape Forecasting}\label{sec:DLF}
In this part, we formulate our deep learning method with censorship handling for bid landscape forecasting, which we call as Deep Landscape Forecasting (DLF) model.

\subsubsection{Discrete Price Model}
First we transform the modeling from continuous space to discrete space.
Note that, since all the price in real-time bidding advertising is discrete, it is natural to propose the discrete price model and derive the probability functions in the discrete price schema.

In the discrete context, a set of $L$ prices $0 < b_1 < b_2 <\ldots < b_L$ is obtained which arises from the finite precision of price determinations.
Analogously we may also consider the grouping of continuous prices as $l = 1, 2, \ldots, L$ uniformly divided disjoint intervals $V_l = (b_l, b_{l+1}]$ where $b_0 = 0$ and $b_l$ is the last observation interval boundary for the given sample, i.e., the proposed bid price in the auction.
$V_L$ is the largest price interval in the whole price space.
This setting is appropriately suited in our task and has been widely used in medical research \cite{Li2016mtlsa} and bid landscape forecasting field \cite{zhang2016bidaware,wang2016functional} where the price is always integer \cite{zhang2014real} thus we set $b_{l+1} - b_l = 1$.

As such, our winning function and losing function over discrete price space is
\begin{equation}\label{eq:win-lose-def}
\begin{aligned}
W(b_l) & \doteq \text{Pr}(z < b_l) = \sum_{j < l} \text{Pr}(z \in V_j) ~, \\
S(b_l) & \doteq \text{Pr}(b_l \leq z) = \sum_{j \geq l} \text{Pr}(z \in V_j) ~,
\end{aligned}
\end{equation}
where the input to the two functions is the bid price $b_l$ from the advertiser.
And the discrete market price probability function at the $l$-th price interval is
\begin{equation}\label{eq:discrete-pdf-def}
\begin{aligned}
p_l &= \text{Pr}(z \in V_l) = W(b_{l+1}) - W(b_l) \\
&= \left[ 1 - S(b_{l+1}) \right] - \left[ 1 - S(b_l) \right]  \\
&= S(b_l) - S(b_{l+1}) ~.
\end{aligned}
\end{equation}

We define the \textit{conditional winning probability} given the price $b_l$ as
\begin{equation}\label{eq:discrete-instant-win-func}
h_l = \text{Pr}(z \in V_l | z \geq b_{l-1}) = \frac{\text{Pr}(z \in V_l)}{\text{Pr}(z \geq b_{l-1})} = \frac{p_l}{S(b_{l-1})} ~,
\end{equation}
which means the probability that the market price $z$ lies in the interval $V_l = (b_l, b_{l+1}]$ given the condition that $z$ is larger than the bid prices which are smaller than $b_l$.
The meaning of $h_l$ is the conditional probability of \textit{just} winning the auction by proposing the bid price at the $l$-th price interval.

\subsubsection{Recurrent Neural Network Model}
Till now, we have presented the discrete price model and discuss the winning and losing probability over the discrete price space.
We here propose our DLF model based on recurrent neural network $f_{\btheta}$ with the parameter $\btheta$, which captures the sequential patterns for conditional probability $h^i_l$ at every price interval $V_l$ for the $i$-th sample.

\begin{figure}[h]
  \centering
  \includegraphics[width=1.0\columnwidth]{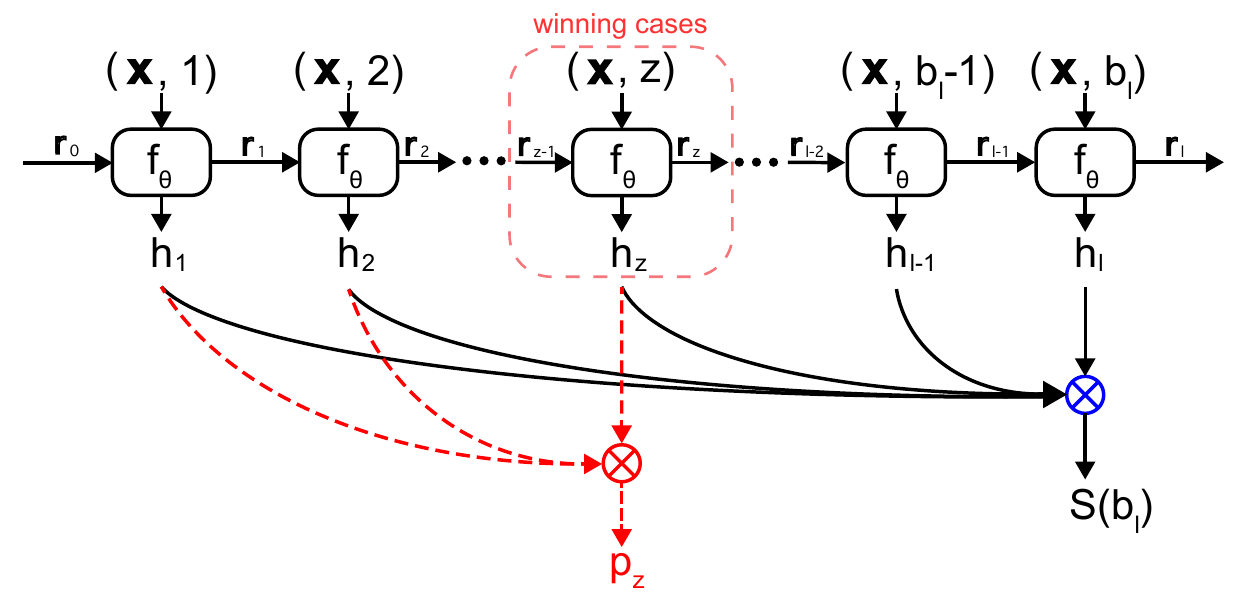}
  \caption{Detailed illustration of DLF model. Note that only the winning cases have the true market price and can calculate $p_z$ for the loss $L_1$. 
  }\label{fig:DLF}
\end{figure}

The detailed structure of DLF network is illustrated in Figure~\ref{fig:DLF}.
At each price interval $V_l$, the $l$-th RNN cell predicts the conditional winning probability $h^i_l$ given the bid request feature $\bx^i$ conditioned upon the previous events as
\begin{equation}\label{eq:discrete-hazard-function}
\begin{aligned}
h^i_l &= \text{Pr}(z \in V_l ~|~ z \geq b_{l-1} ,~ \bx^i; \btheta) \\
&= f_{\btheta}(\bx^i, b_l ~|~ \br_{l-1}) ~,
\end{aligned}
\end{equation}
where $f_{\btheta}$ is the RNN function taking $(\bx^i, b_l)$ as input and $h^i_l$ as output.
$\br_{l-1}$ is the hidden vector calculated from the last RNN cell.
In our paper we implement the RNN function as a standard LSTM unit \cite{hochreiter1997long}, which has been widely used in sequence data modeling. We describe the implementation details of $f_{\btheta}$ in the appendix.

From Eqs.~(\ref{eq:win-lose-def}), (\ref{eq:discrete-instant-win-func}) and (\ref{eq:discrete-hazard-function}), we can easily derive the losing probability function $S(b)$ and the winning probability function $W(b)$ with the bidding price $b$ for the $i$-th individual sample as
\begin{equation}\label{eq:discrete-win-survival-def}
\begin{aligned}
S(b | \bx^i; \btheta) &= \text{Pr}(b \leq z | \bx^i; \btheta) = \text{Pr}(z \not\in V_1, z \not\in V_2, \dots, z \not\in V_{l} | \bx^i; \btheta) \\
&= \text{Pr}(z \not\in V_1 |\bx^i; \btheta) \cdot \text{Pr}(z \not\in V_2 | z \not\in V_1, \bx^i; \btheta) \cdots \\
& ~~~~~~~~~~~~~~~~\cdot \text{Pr}(z \not\in V_{l_b} | z \not\in V_1, \ldots, z \not\in V_{l_b-1}, \bx^i; \btheta) \\
& = \prod_{l: l \leq l_b} \left[ 1- \text{Pr}(z \in V_l ~|~ z \geq b_l  ,~ \bx^i; \btheta) \right] = \prod_{l: l \leq l_b} (1 - h^i_l) ~, \\
W(b | \bx^i; \btheta) &= \text{Pr}(b > z  | \bx^i; \btheta) = 1 - S(b | \bx^i; \btheta) = 1- \prod_{l: l \leq l_b} (1 - h^i_l)~,
\end{aligned}
\end{equation}
where $l_b$ is the price interval index for $b$.
Here we use probability chain rule to calculate the joint losing probability $S(b)$ at the given bid price $b$ through multiplying the conditional losing probability $(1-h_l)$, i.e., inverse of the conditional winning probability.

Moreover, taking Eqs.~(\ref{eq:discrete-pdf-def}) and (\ref{eq:discrete-instant-win-func}) into consideration, the probability of market price $z$ directly lying in the interval of $V_l$ for the $i$-th sample is
\begin{equation}\label{eq:pdf-calc}
p^i_l = \text{Pr}(z^i \in V_l | \bx^i; \btheta) = h^i_{l^i} \prod_{l: l < l^i} (1 - h^i_l) ~.
\end{equation}

\subsubsection{Loss Functions}
Since there is no ground truth of either market price distribution or winning probability, here we maximize the log-likelihood over the empirical data distribution to learn our deep model. 
We consider from two aspects for the loss function.

The first loss is based on the P.D.F. and it aims to minimize the negative log-likelihood of the market price $z = z^i$ over the winning logs as
\begin{equation}\label{eq:market-price-objective}
\begin{aligned}
L_1 &= - \log \prod_{(\bx^i, z^i) \in \mathbb{D}_{\text{win}}} \text{Pr}(z^i \in V_{l^i} | \bx^i; \btheta) = - \log \prod_{(\bx^i, z^i) \in \mathbb{D}_{\text{win}}} p^i_l \\
&= - \log \prod_{(\bx^i, z^i) \in \mathbb{D}_{\text{win}}} h^i_{l^i} \prod_{l: l < l^i} (1 - h^i_l) \\
&= - \sum_{(\bx^i, z^i) \in \mathbb{D}_{\text{win}}} \left[ \log h^i_{l^i}  + \sum_{l: l < l^i} \log(1 - h^i_l)  \right] ~,
\end{aligned}
\end{equation}
where $l^i$ is the interval index of the true market price $z^i \in V_{l^i}$ given the feature vector $\bx^i$.

\begin{figure}[h]
  \centering
  \includegraphics[width=0.9\columnwidth]{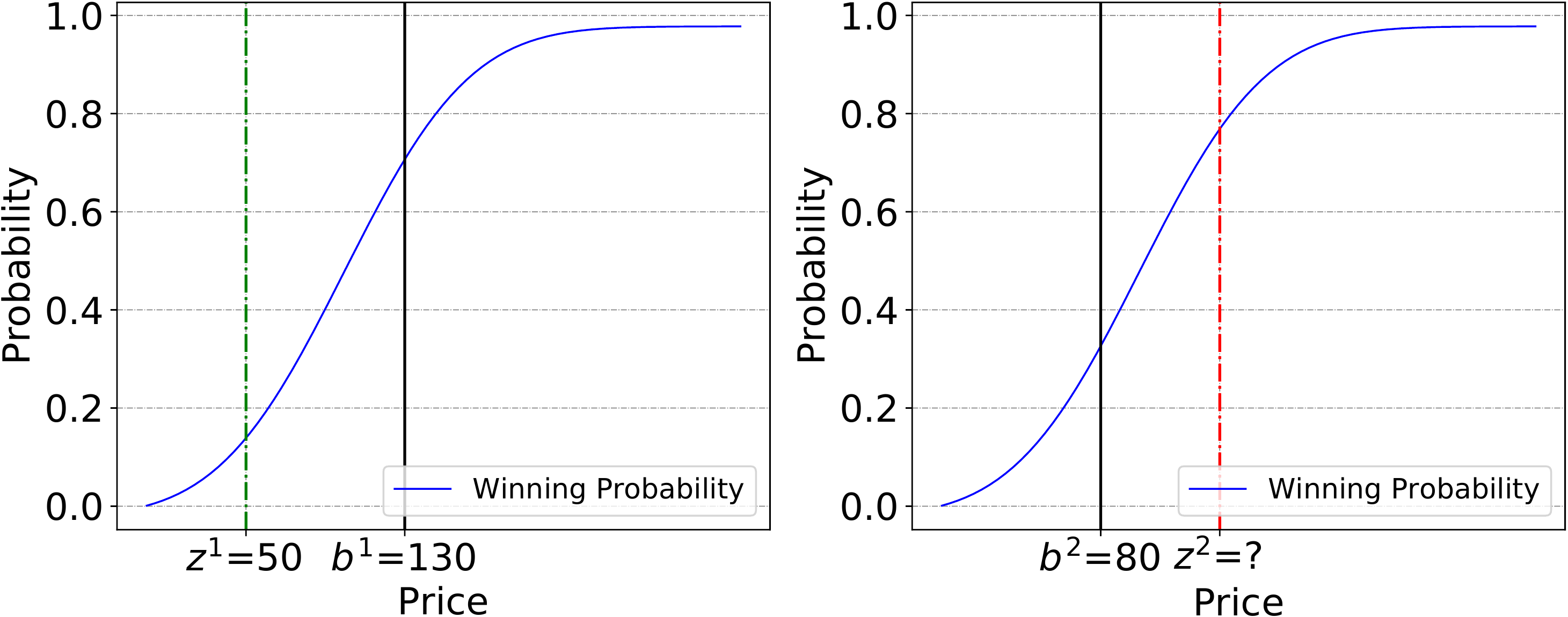}
  \caption{Two examples of winning curves.}\label{fig:win-curve}
\end{figure}
The second loss is based on the C.D.F., i.e., winning probability distribution.
Recall that there are winning cases and losing cases in the dataset.
As is shown in Figure~\ref{fig:win-curve}, the left subfigure is the winning case where $z^1$ has been known and $z^1 < b^1$; The right figure is the losing case where $z^2$ is unknown (censored) but we only have the knowledge that $z^2 \geq b^2$.
Thus, there are two motivations about the second loss.

For the winning cases as in the left part of Figure~\ref{fig:win-curve}, we need to ``push down'' the winning probability during the price range of $[0, z]$, while ``pull up'' the winning probability during the price range of $[z, \infty)$, especially for the winning probability in $[z,b]$.
Thus, on one hand, we adopt the loss over the winning cases that
\begin{equation}\label{eq:winning-objective}
\begin{aligned}
L_{\text{win}} &= - \log \prod_{(\bx^i, b^i) \in \mathbb{D}_{\text{win}}} \text{Pr}(b^i > z | \bx^i; \btheta) \\
&= - \log \prod_{(\bx^i, b^i) \in \mathbb{D}_{\text{win}}} W(b^i | \bx^i; \btheta) \\
&= - \sum_{(\bx^i, b^i) \in \mathbb{D}_{\text{win}}} \log \Big[ 1- \prod_{l: l \leq l^i} (1 - h^i_l) \Big] ~.
\end{aligned}
\end{equation}

As for the the losing cases in the right part of Figure~\ref{fig:win-curve}, we just need to ``push down'' the winning probability since we have no idea about the true market price but we only know that $z \geq b$.
On the other hand, we just adopt the loss over the losing auctions as
\begin{equation}\label{eq:losing-objective}
\begin{aligned}
L_{\text{lose}} &= - \log \prod_{(\bx^i, b^i) \in \mathbb{D_{\text{lose}}}} \text{Pr}(z \geq b^i | \bx^i; \btheta) \\
&= - \log \prod_{(\bx^i, b^i) \in \mathbb{D_{\text{lose}}}} S(b^i | \bx^i; \btheta) \\
&= - \sum_{(\bx^i, b^i) \in \mathbb{D_{\text{lose}}}} \sum_{l: l \leq l^i} \log (1 - h^i_l) ~.
\end{aligned}
\end{equation}

\subsection{Model Realization}\label{sec:model-realization}
In this section, we unscramble some intrinsic properties of our deep model and analyze the model efficiency in this section.

\minisection{Properties of Loss Function} First of all, we take the view of winning prediction of our methodology.
As is known that there is a winning status, i.e., an indicator of winning the auction, for each sample as
\begin{equation}\label{eq:lose-indicator}
w^i=
\begin{cases}
1, & \text{if}~~~~~~~~~~ b^i > z^i ~,
\cr 0, & \text{otherwise}~~ b^i \leq z^i ~.
\end{cases}
\end{equation}
For the winning logs, each sample $(\bx^i, z^i, b^i)$ is uncensored (i.e., $z^i$ is known) where $w^i = 1$.
While for the losing logs, the true market price $z^i$ is unknown but the advertiser only has the idea that $z^i \geq b^i$, so that $w^i = 0$.

Therefore, taking Eqs.~(\ref{eq:winning-objective}) and (\ref{eq:losing-objective}) altogether and we may find that the combination of $L_{\text{win}}$ and $L_{\text{lose}}$ describes the classification of winning the auction as
\begin{align}
& L_{2} = L_{\text{win}} + L_{\text{lose}} \nonumber \\
&= - \log \prod_{(\bx^i, b^i) \in \mathbb{D}_{\text{win}}} \text{Pr}(b^i > z | \bx^i; \btheta) \nonumber  - \log \prod_{(\bx^i, b^i) \in \mathbb{D_{\text{lose}}}} \text{Pr}(z \geq b^i | \bx^i; \btheta) \nonumber \\
&= - \log \prod_{(\bx^i, b^i) \in \mathbb{D}} \left[ W(b^i | \bx^i; \btheta) \right]^{w^i} \cdot \left[ 1 - W(b^i | \bx^i; \btheta) \right]^{1-w^i} \label{eq:classification-loss} \\
&= - \sum_{(\bx^i, b^i) \in \mathbb{D}} \left\{ w^i \cdot \log W(b^i | \bx^i; \btheta) \right. \nonumber + \left. (1-w^i) \log \left[ 1 - W(b^i | \bx^i; \btheta) \right] \right\} ~, \nonumber
\end{align}
which is the cross entropy loss for predicting winning probability when bidding at $b^i$ given $\bx^i$ over all the data $\mathbb{D} = \mathbb{D}_{\text{win}} \bigcup \mathbb{D}_{\text{lose}}$.

Combining all the objective functions and our goal is to minimize the negative log-likelihood over all the data samples including both winning logs and losing logs as
\begin{equation}\label{eq:total-loss}
\arg \min_{\btheta}~~ \alpha L_1 + (1-\alpha) L_{2} ~,
\end{equation}
where the hyperparameter $\alpha$ controls the order of magnitudes of the gradients from the two losses at the same level to stabilize the model training.

In the traditional survival analysis methods \cite{cox1992regression,katzman2018deepsurv} and the related works for bid landscape forecasting \cite{wu2015predicting,zhu2017gamma}, they usually adopts only $L_1$ based on P.D.F. and $L_{\text{lose}}$ for censorship handling.
We propose a comprehensive loss function which learns from both winning logs and losing logs.
From the discussion above, $L_{\text{win}}$ and $L_{\text{lose}}$ collaboratively learns the data distribution from the C.D.F. view.

\minisection{Model Efficiency}
Here we analyze the computational complexity of our DLF model.
As is shown in Eq.~(\ref{eq:discrete-hazard-function}), each recurrent unit $f_{\btheta}$ takes ($\bx$, $b_l$, $\br_{l-1}$) as input and outputs probability scalar $h_l$ and hidden vector $\br_l$ to the next unit.
Recall that the maximal price interval is $L$, so the calculation of the recurrent units will runs for maximal $L$ times.
We assume the average case time performance of recurrent units $f_{\btheta}$ is $O(C)$, which is related to the implementation of the unit \cite{zhang2016architectural}, e.g., recurrent depth, recurrent skip coefficients, yet can be parallelized through GPU processor.
The subsequent calculation is to obtain the multiplication results of $h_l$ or ($1-h_l$) to get the results of $p(z)$ and $S(b)$, as that in Figure~\ref{fig:DLF}, whose complexity is $O(L)$.
Thus the overall time complexity of DLF model is $O(CL)+O(L) = O(CL)$, which is the same as the original recurrent neural network model.

In many literatures, recurrent neural networks have been deployed in recommender system \cite{wu2016personal}, online advertising platform \cite{zhou2018deepb} and machine translation system \cite{wu2016google}, each of which shows promising time efficiency in large scale online systems and, to some extent, guarantees online inference efficiency for our DLF model.
We may also optimize the implementation of the RNN unit through other techniques, such as Quasi-RNN \cite{bradbury2016quasi} and sliced-RNN \cite{yu2018sliced}.
Moreover, the landscape forecasting module could be parallelly executed with the utility estimation in RTB scenario, e.g., click-through rate prediction model, and jointly feed the results for final bid decision making.
In our experiments, under the recommended settings, we evaluate our model and it achieved 22 milliseconds for averaged inference time given one sample, satisfying the 100 milliseconds requirement of bid decision in the RTB scenario \cite{wang2017display}.

\section{Experiments}\label{sec:exp}
In this section, we present the experimental setup and the corresponding results under various evaluation metrics with significance test.
Furthermore, we look deeper into our model and analyze some insights of the experiment results.
Moreover, we have published the implementation code for reproducible experiments\footnote{Reproducible code: https://github.com/rk2900/DLF.}.

\subsection{Datasets and Experiment Flow}
We use two real-world RTB datasets in our experiment.
\textbf{iPinYou} RTB dataset, which has been published in \cite{liao2014ipinyou},
contains 64.7M bidding records, 19.5M impressions, 14.79K clicks and 16.0K CNY expense on 9 campaigns from different advertisers during 10 days in 2013. Each bidding log has 16 attributes, including weekday, hour, user agent, region, ad slot ID, etc.
The auctions during the last 3 days are set as test data while the rest as training data.
The other bidding dataset is \textbf{YOYI} dataset which was published in \cite{ren2016user}.
It includes 402M impressions, 500K clicks and 428K CNY expense during 8 days in Jan. 2016.
More details of the two datasets have been provided in \cite{liao2014ipinyou} and \cite{ren2016user} respectively.

\minisection{Data Preparation} For simulating the real bidding market in an online fashion and show the advantages of our deep survival model, we take the original data of impression log as full-volume auction data, and perform a truthful bidding strategy \cite{lee2012estimating} to simulate the bidding process, which produces the winning bid dataset $\mathbb{D}_{\text{win}}$ and the losing bid dataset $\mathbb{D}_{\text{lose}}$ respectively.
For each data sample $(\bx_{\text{win}}, b_{\text{win}}, z_{\text{win}}) \in \mathbb{D}_{\text{win}}$, the real market price $z_{\text{win}} < b_{\text{win}}$ is known for the advertisers, while for each $(\bx_{\text{lose}}, b_{\text{lose}}, z_{\text{lose}}) \in \mathbb{D}_{\text{lose}}$ the corresponding market price $z_{\text{lose}} \geq b_{\text{lose}}$ is hidden.
It guarantees the similar situation as that faced by all the advertisers in the real world marketplace.
This simulation and data processing method have been widely used in bid landscape forecasting literatures \cite{wu2015predicting,zhu2017gamma,zhang2016bidaware,wang2016functional}.

After data preparation, we make some statistics over the resulted datasets.
As is illustrated in Table~\ref{tab:statistics}, we can find that the averaged market price in $\mathbb{D}_{\text{win}}$ is much lower than that of $\mathbb{D}_{\text{lose}}$, which is reasonable because of the second-price auction mechanism and also reflects the bias of the model without using losing (censored) logs.

In these datasets, since all the prices are integer value, we bucketize the discrete price interval as $\text{interval\_size}=1$ and the maximal price interval number $L$ is equal to the largest integer price in the dataset where $L=300$.

\begin{table}[t]
  \centering
  \caption{The statistics of the datasets. WR: winning rate; AMP: averaged market price.}\label{tab:statistics}
  \resizebox{\columnwidth}{!}{
    \begin{tabular}{c|c|c|c|c|c|c}
      \hline
      Campaign & Total \# & Winning \# & ~~WR~~ & ~~AMP~~ & AMP ($\mathbb{D}_{\text{win}}$) & AMP ($\mathbb{D}_{\text{lose}}$) \\
      \hline
      1458 & 3,697,694 & 1,116,644 & 0.3020 & 69.6696 & 27.4265 & 87.9452 \\
      2259 & 1,252,753 & 396,283 & 0.3163 & 96.7888 & 27.1986 & 128.9877 \\
      2261 & 1,031,479 & 321,931 & 0.3121 & 87.6479 & 18.9000 & 118.8396 \\
      2821 & 1,984,525 & 228,833 & 0.1153 & 93.8962 & 13.2118 & 104.4125 \\
      2997 & 468,500 & 70,747 & 0.1510 & 60.4188 & 7.2762 & 69.8711 \\
      3358 & 2,043,032 & 315,010 & 0.1542 & 95.4967 & 21.2540 & 109.0308 \\
      3386 & 3,393,223 & 819,447 & 0.2415 & 78.0327 & 23.8983 & 95.2682 \\
      3427 & 3,130,560 & 654,989 & 0.2092 & 81.9650 & 25.2118 & 96.9808 \\
      3476 & 2,494,208 & 723,847 & 0.2902 & 80.0719 & 31.2218 & 100.0453 \\
      \hline
      Overall & 19495974 & 4647731 & 0.2384 & 82.0744 & 25.0484 & 99.9244 \\
      \hline
      \hline
      YOYI & 401,617,064 & 202,214,191 & 0.5035 & 55.7444 & 24.4488 & 87.4842 \\
      \hline
    \end{tabular}
  }
\end{table}

\minisection{Evaluation Phase}
In this phase, the corresponding market price distribution $p(z | \bx)$ with the true market price of each sample $\bx$ in the test data is estimated by all of the compared models respectively.
The corresponding winning function $W(b | \bx)$ and losing function $S(b | \bx)$ can be easily obtained through the forecasted market price distribution $p(z | \bx)$ as that in Eqs.~(\ref{eq:win-prob-definition}), (\ref{eq:survival-definition}) and (\ref{eq:win-lose-def}).
We assess the performance of different settings in several measurements, as listed in the next subsection.

\subsection{Evaluation Metrics}\label{sec:eval_metrics}
In our experiments, we evaluate all the models under two metrics and conduct the significance test between our model and the other baselines. Note that, there are two goals for bid landscape forecasting, i.e., forecasting of market price distribution (P.D.F.) and the corresponding winning probability (C.D.F.) estimation given arbitrary bid prices.

First we use average negative log probability (\textbf{ANLP}) as \cite{wang2016functional} to evaluate the performance of forecasting the market price distribution.
Specifically, ANLP is to assess the likelihood of the co-occurrence of the test bid requests with the corresponding market prices, which is calculated as
\begin{equation}
\vspace{-5pt}
\bar{P} = - \frac{1}{|\mathbb{D}_{\text{test}}|} \sum_{(\bx^i, z^i) \in \mathbb{D}_{\text{test}}} \log p_z(z^i|\bx^i) ~,
\end{equation}
where $p_z(z|\bx)$ is the learned bid landscape forecasting function of each model.

The last evaluation metric is concordance index (\textbf{C-index}), which is the most common evaluation used in survival analysis \cite{harrell1984regression,Li2016mtlsa,luck2017deep} and reflects a measure of how well a model predicts the ordering of samples according to their market prices.
That is, given the bid price $b$, two auction samples $d^1 = (\bx^1, z^1)$ with large market price $z^1 \geq b$ and $d^2 = (\bx^2, z^2)$ with small market price $z^2 < b$ should be ordered as $d^1 \prec d^2$ where $d^1$ is placed before $d^2$.
This evaluation is the same as the area under ROC curve (AUC) metric in the classification tasks \cite{qu2016product,ren2016user} when there is only one event of interest (i.e., winning in our task) \cite{Li2016mtlsa}.
From the classification view of auction winning probability estimation by proposing $b$, C-index assesses the ordering performance among all the winning and losing pairs at $b$ among the test data thus illustrates the performance of winning probability estimation.

Finally, we conduct the significance test to verify the statistical significance of the performance improvement of our model w.r.t. the baseline models.
Specifically, we deploy a Mann-Whitney U test \cite{mason2002areas} under C-index metric, and a t-test \cite{bhattacharya2002median} under ANLP metric.


\subsection{Compared Settings}\label{sec:campared-settings}
We compare our proposed DLF model with nine baseline models including traditional Cox proportional hazard function model, survival tree model, multi-task learning method and other deep learning models.

\begin{itemize}[leftmargin=4mm]
  \item \textbf{KM} is Kaplan-Meier estimator \cite{kaplan1958nonparametric} which is a statistic-based non-parametric method and has been used in several bid landscape forecasting works \cite{zhang2016bidaware,wang2016functional}.
  \item \textbf{Lasso-Cox} is a semi-parametric method \cite{zhang2007adaptive} based on Cox proportional hazard model \cite{cox1992regression} with $l1$-regularization.
  \item \textbf{DeepSurv} is a Cox proportional hazard model with deep neural network \cite{katzman2018deepsurv} for feature extraction upon the sample covariates. The loss function is the negative partial likelihood of the winning and losing outcomes.
  \item \textbf{Gamma} is the gamma distribution based regression model \cite{zhu2017gamma}. The winning price of each bid request is modeled by a unique gamma distribution with respect to its features.
  \item \textbf{MM} is the mixture regression model. This model uses both linear regression and censored regression, and combines two models and predicts as a mixture manner \cite{wu2015predicting}. 
  \item \textbf{MTLSA} is the recently proposed multi-task learning model \cite{Li2016mtlsa}. It transforms the original survival analysis problem into a series of binary classification problems, and uses a multi-task learning method. The original model predicts the death rate of a patient, we change it to predict the wining rate of bidding in an auction.
  \item \textbf{STM} is the survival tree model. This model combines Kaplan-Meier estimator and decision trees with bi-clustering to predict bid landscape. This model is proposed in \cite{wang2016functional} and, to our knowledge, achieved state-of-the-art performance in bid landscape forecasting.
  \item \textbf{DeepHit} is a deep neural network model \cite{lee2018deephit} which predicts the probability of each bidding price from the minimum price to the maximum price.
  \item \textbf{DWPP} \cite{wu2018deep} is a deep winning price prediction method using neural network to directly predict the market price, with assumption of the distribution form as Gaussian distribution.
  \item \textbf{RNN} is based on our DLF model. However, it only optimizes over the winning logs \textit{without} considering the censored information, whose loss function is only $(L_1+L_{\text{win}})$. This model is used to illustrate the power of partial likelihood loss $L_{\text{lose}}$ over the censored losing data.
  \item \textbf{DLF} is our deep landscape forecasting model which has been described in Section~\ref{sec:method}.
\end{itemize}
The details of the experimental configurations, such as hardware and training procedure, have been included in the appendix.

\subsection{Experimental Results}
In this part, we present the detailed performance of all the compared models over the evaluation metrics.

\begin{table}[h]
  \centering
  \caption{Averaged negative log probability performance: the smaller, the better. (* indicates p-value $<10^{-6}$ in significance test).}\label{tab:anlp}
  \resizebox{\columnwidth}{!}{
    \begin{tabular}{c|ccccccccccc}
      \hline
      & \multicolumn{11}{c}{ANLP} \\
      iPinYou & KM & Lasso-Cox & DeepSurv & Gamma & MM & MTLSA & STM & DeepHit & DWPP & RNN & DLF \\
      \hline
      1458 & 10.532 & 38.608 & 38.652 & 5.956 & 5.788 & 9.791 & 4.761 & 5.510 & 29.204 & 9.506 & \bf 4.088\textsuperscript{*}\\
      2259 & 14.671 & 28.234 & 29.658 & 6.069 & 7.328 & 10.248 & 5.471 & 5.586 & 39.263 & 9.625 & \bf 5.244\textsuperscript{*}\\
      2261 & 14.665 & 39.129 & 39.390 & 5.986 & 7.020 & 10.261 & 4.818 & 5.442 & 32.805 & 9.417 & \bf 4.632\textsuperscript{*}\\
      2821 & 19.582 & 43.099 & 43.072 & 7.838 & 7.262 & 9.895 & 5.572 & 5.614 & 40.537 & 23.099 & \bf 5.428\textsuperscript{*}\\
      2997 & 16.203 & 32.849 & 33.052 & 5.999 & 6.702 & 9.167 & 5.083 & 5.470 & 34.940 & 16.639 & \bf 4.504\textsuperscript{*}\\
      3358 & 19.253 & 44.769 & 44.885 & 6.736 & 7.177 & 9.484 & 5.539 & 5.616 & 40.958 & 13.806 & \bf 5.281\textsuperscript{*}\\
      3386 & 15.973 & 39.781 & 41.943 & 6.488 & 6.141 & 8.834 & 5.228 & 5.549 & 32.550 & 10.743 & \bf 4.940\textsuperscript{*}\\
      3427 & 16.902 & 41.558 & 41.698 & 6.002 & 6.185 & 9.090 & 5.321 & 5.552 & 33.387 & 9.565 & \bf 4.836\textsuperscript{*}\\
      3476 & 10.507 & 39.551 & 39.518 & 5.710 & 6.022 & 10.240 & 4.537 & 5.554 & 31.609 & 7.891 & \bf 4.012\textsuperscript{*}\\
      \hline
      Overall & 15.366 & 38.620 & 39.096 & 6.310 & 6.552 & 9.668 & 5.148 & 5.544 & 35.028 &12.255 & \bf 4.774\textsuperscript{*}\\
      \hline
      \hline
      YOYI & 7.907 & 30.946 & 27.897 & 6.475 & 5.652 & 10.286 & 4.503 & 5.567 & 29.108 & 5.885 & \bf 4.453\textsuperscript{*}\\
      \hline
    \end{tabular}
  }
    \vspace{-10pt}
\end{table}

\subsubsection{Landscape Forecasting Performance}
We first analyze the probability density estimation performance for learning the bid landscape forecasting.
Though there is no ground truth for the market price distribution $p(z)$, we may also use the negative log-likelihood result to evaluate the performance over the test data.
Table~\ref{tab:anlp} lists the ANLP performance of the compared models.
From the table, we may find that our DLF model achieves significant improvements against the other baselines 
including the state-of-the-art model STM on both iPinYou and YOYI datasets.

We also find from the table that
(i) The survival tree model STM achieves relatively better performance than other baselines which may be the result of the well clustering methodology and the non-parametric survival analysis.
(ii) All of the models with survival analysis, i.e., DeepSurv, Gamma, MM, STM, MTLSA, DeepHit, perform much better than RNN model which does not consider censored data into model training.
(iii) DeepHit model gets worse results than DLF model probably for the reason that it does not model the price-level sequential dependency as that in our method, which in contrast reflects that the conditional sequential modeling of DLF model in Eq.~(\ref{eq:discrete-hazard-function}) has significantly improved the forecasting performance in the market price distribution modeling.
(iv) Though DWPP utilizes deep model for feature extraction, it performs poor under ANLP metric which has also been reported in their paper \cite{wu2018deep}. The reason may be the assumed Gaussian form of the market price distribution lacks generalization in the practical applications.

\begin{table}[h]
  \centering
  \caption{C-index performance: the larger, the better, (* indicates p-value $<10^{-6}$ in significance test).}\label{tab:c-index}
  \resizebox{\columnwidth}{!}{
    \begin{tabular}{c|ccccccccccc}
      \hline
      & \multicolumn{11}{c}{C-index} \\
      iPinYou & KM & Lasso-Cox & DeepSurv & Gamma & MM & MTLSA & STM & DeepHit & DWPP & RNN & DLF \\
      \hline
      1458 & 0.698 & 0.820 & 0.835 & 0.612 & 0.698 & 0.505 & 0.764 & 0.861 & 0.866 & 0.894 & \bf 0.904\textsuperscript{*}\\
      2259 & 0.685 & 0.775 & 0.791 & 0.584 & 0.685 & 0.505 & 0.768 & 0.785 & 0.729 & 0.791 & \bf 0.876\textsuperscript{*}\\ 
      2261 & 0.666 & 0.847 & 0.890 & 0.564 & 0.666 & 0.508 & 0.812 & 0.838 & 0.807 & 0.874 & \bf 0.929\textsuperscript{*}\\
      2821 & 0.677 & 0.741 & 0.714 & 0.563 & 0.678 & 0.507 & 0.790 & 0.810 & 0.746 & 0.737 & \bf 0.881\textsuperscript{*}\\
      2997 & 0.734 & 0.910 & 0.852 & 0.641 & 0.734 & 0.517 & 0.835 & 0.907 & 0.885 & 0.762 & \bf 0.919\textsuperscript{*}\\
      3358 & 0.704 & 0.866 & 0.896 & 0.601 & 0.706 & 0.542 & 0.811 & 0.888 & 0.744 & 0.819 & \bf 0.944\textsuperscript{*}\\
      3386 & 0.716 & 0.845 & 0.854 & 0.569 & 0.719 & 0.512 & 0.849 & 0.881 & 0.833 & 0.800 & \bf 0.923\textsuperscript{*}\\
      3427 & 0.724 & 0.830 & 0.845 & 0.586 & 0.742 & 0.508 & 0.798 & 0.873 & 0.796 & 0.804 & \bf 0.901\textsuperscript{*}\\
      3476 & 0.692 & 0.865 & 0.877 & 0.676 & 0.692 & 0.505 & 0.830 & 0.879 & 0.861 & 0.917 & \bf 0.922\textsuperscript{*}\\
      \hline
      Overall & 0.700 & 0.834 & 0.840 & 0.600 & 0.703 & 0.513 & 0.807 & 0.858 & 0.807 & 0.823 & \bf 0.911\textsuperscript{*}\\
      \hline
      \hline
      YOYI & 0.791 & 0.847 & 0.862 & 0.528 & 0.791 & 0.510 & 0.886 & 0.878 & 0.856 & 0.898 & \bf 0.924\textsuperscript{*} \\
      \hline
    \end{tabular}
  }
    \vspace{-10pt}
\end{table}
\subsubsection{Winning Prediction Performance}
In this part we illustrate the measurement of the winning prediction under the given bid price $b$ of the sample $\bx$.
As is discussed before, this can be regarded as a binary classification problem, so we present the performance of C-index in Table~\ref{tab:c-index}.
From the table we can observe that DLF achieves the best C-index value among all the compared models on both iPinYou and YOYI datasets which show the classification effectiveness of our model.
Especially, our model gains over 12.9\%  average improvements against the state-of-the-art STM model.

We can also conduct below findings from the table.
(i) All the deep models including DeepSurv, DeepHit, DWPP, RNN and DLF have relatively better C-index performance than the other non-deep baselines.
Even the RNN model without any censorship handling achieves satisfying prediction performance, which again reflects the advantage of our novel modeling perspective with sequential pattern mining.
(ii) MM and Gamma do not perform well which may be accounted for that these models adopt restricted assumptions of the base distribution for the probability dense function. This phenomenon verifies our analysis in Sections.~\ref{sec:intro} and \ref{sec:related-work} and reveals the importance of modeling without distributional assumptions.
(iii) DWPP performs not well which is reasonable because it is optimized for market price regression rather than winning probability estimation.
However, please note that, the forecasting of market price distribution and the corresponding winning probability estimation are more general for RTB advertising.

\subsubsection{Model Convergence}
\begin{figure}[h]
  \centering
  \includegraphics[width=0.8\columnwidth]{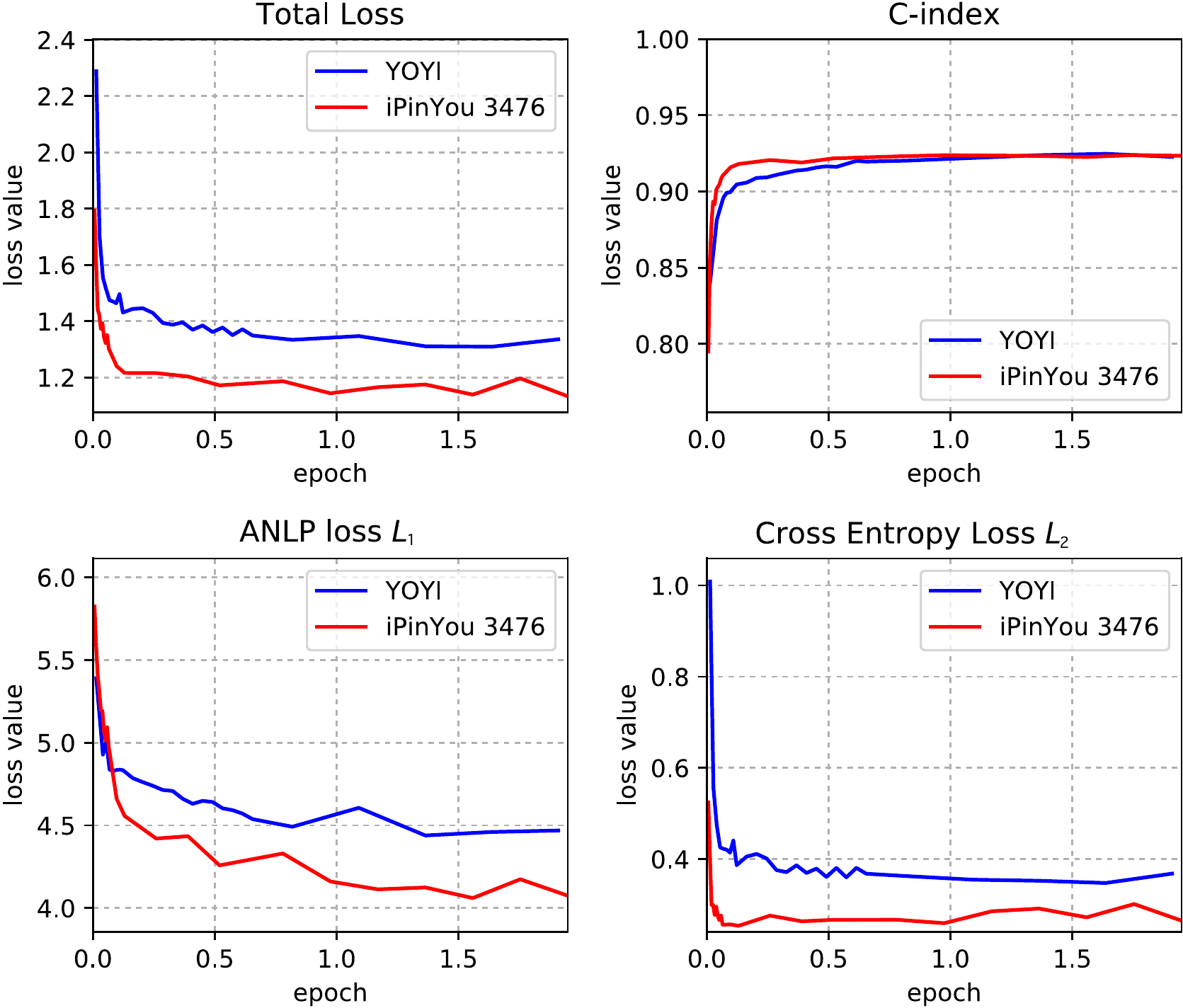}
  \caption{Learning curves over Campaign 3476 of iPinYou and YOYI. Here ``epoch'' means one iteration over the whole training data and $\alpha=0.25$ in Eq.~(\ref{eq:total-loss}).}\label{fig:learn-curve}
  \vspace{-12pt}
\end{figure}

To illustrate the model training and convergence of DLF model, we plot the learning curve and the evaluation results on iPinYou Campaign 3476 and YOYI datasets in Figure~\ref{fig:learn-curve}.
Recall that our model optimizes over two loss functions, i.e., the ANLP loss $L_1$ and the cross entropy loss $L_{2}$.
We apply a method \cite{cao2018neural} whose idea is to feed the batch of training data under each one of the two losses alternatively.
From the learning curves we can find that 
(i) DLF converges quickly and the values of both losses drop to stable convergence at about the first complete iteration over the whole training dataset.
(ii) The two losses are alternatively optimizing and facilitate each other during the  training, which proves the learning stability of our model.

\begin{figure*}[t]
  \centering
  \includegraphics[width=0.6\textwidth]{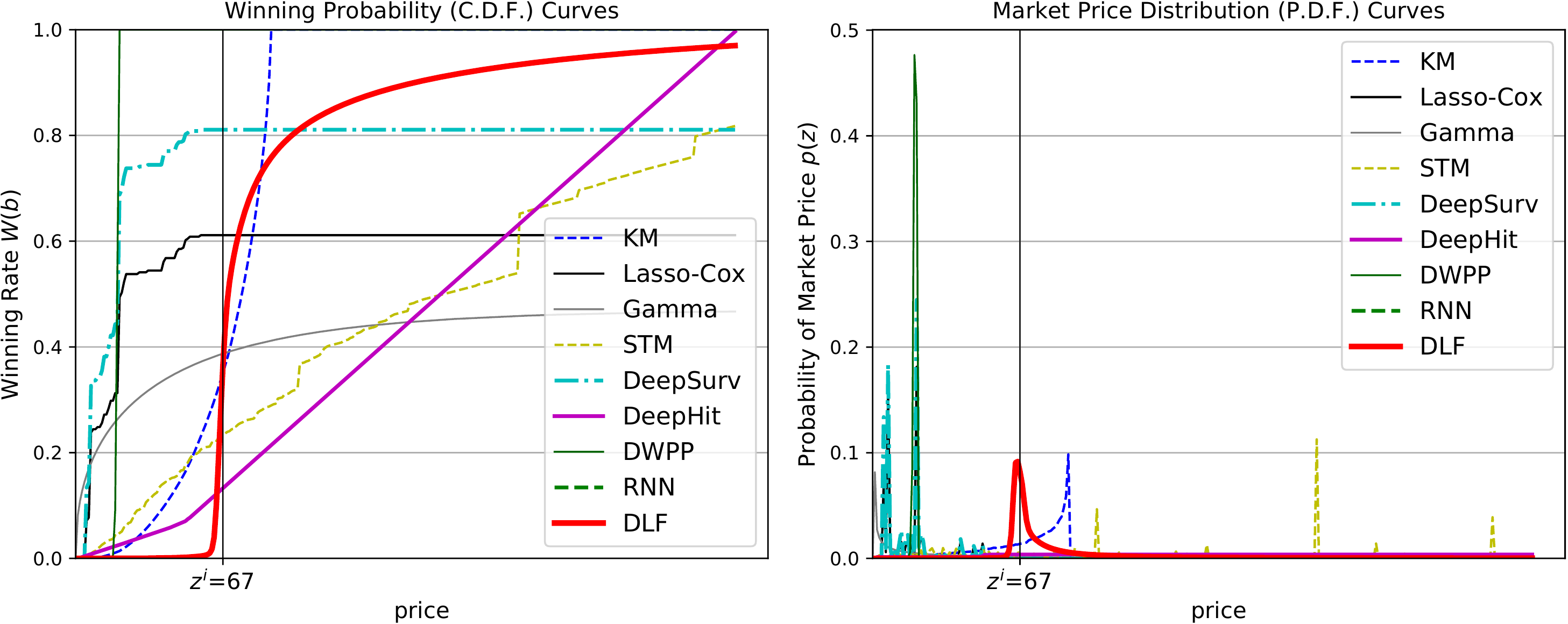}
  \caption{A comprehensive visualization of winning probability $W(b|\bx^i)$ estimation and market price probability $p(z|\bx^i)$ prediction over different models. The vertical black line is the true market price $z^i=67$ of this sample.}\label{fig:visualization}
\vspace{-5pt}
\end{figure*}

\subsection{Further Investigation}
In this section, we illustrate some comprehensive analysis of the prediction results of different models.
As is illustrated in Figure~\ref{fig:visualization}, we plot the winning probability (C.D.F.) and the corresponding market price distribution (P.D.F.) of an example sample, calculated from all the compared models.
The true market price of this sample is $z^i=67$.

\minisection{Accurate Bid Landscape Forecasting} As is illustrated in the figure, when the value of market price distribution $p(z^i)$ is high, the corresponding winning probability increases rapidly.
Thus accurately predicting $p(z^i)$ makes the winning probability estimation reasonable.
From the figure, we may find that our DLF model accurately places \textit{highest} probability density on the true market price $z^i=67$, while the other models cannot conduct reasonable forecasting results.
This finding reflects the advantage of our DLF model under ANLP metric since it directly measures the accuracy of bid landscape forecasting.

\minisection{Flexibility of Forecasted Distribution} As has been discussed before, we do not make any assumptions of the distribution form of either market price distribution or winning probability function, thus our DLF model can model sophisticated distribution forms as illustrated in the figure.
However, those models assuming specific distribution forms, i.e., Lasso-Cox, Gamma, DeepSurv and DWPP, cannot reasonably model this practical case since their strong assumptions lack generalization in real-world applications.
Specifically, Lasso-Cox, DeepSurv and DWPP show similar results of P.D.F. and C.D.F. which may be accounted for the same Gaussian distribution form adopted by these models.
Thus it further shows the disadvantage of assuming distribution forms for bid landscape forecasting.

\minisection{Effective Censorship Handling} Comparing the forecasted result of DLF with that of RNN which lacks the censorship handling, we can find that though RNN model predicts the market price distribution and the winning probability with the similar shape to our DLF model, it over-estimates the winning probability and place the probability density of $p(z^i)$ not accurately.
It is easy to explain since without censorship handling, the model may result in biased landscape forecasting results as also shown in \cite{wang2016functional}.

\section{Conclusion and Future Work}\label{sec:conclusion}
In this paper, we analyzed the challenges of bid landscape forecasting and the cons of state-of-the-art methods.
To tackle with these problems, we proposed a fined-grained bid landscape forecasting model based on deep recurrent neural network and survival analysis modeling.
Note that we do not assume any distribution forms for bid landscape forecasting.
Our DLF model not only captures complex patterns in the bid landscape of each bid request, but also considers prediction of the winning status over both winning logs and losing (censored) logs.
The comprehensive experiment results have shown the significant advantages of our model comparing with other strong baselines.
A deep investigation has been performed to verify the robustness and correctness of our model.

For the future work, we plan to incorporate the proposed bid landscape forecasting model into bid optimization for profit maximization \cite{ren2018bid,lin2016combining,DiemertMeynet2017} in real-time bidding advertising.

\minisection{Acknowledgments}
The corresponding author Weinan Zhang thanks the support of National Natural Science Foundation of China (61702327, 61772333, 61632017) and Shanghai Sailing Program (17YF1428200).

\bibliographystyle{ACM-Reference-Format}
\bibliography{rt0873o}

\newpage
\begin{appendices}
\section{Model Implementations of DLF}
In this section, we describe the details of the implemented architecture of the proposed model (DLF).

Recall that, each sample is a triple $(\bx, z, b_l)$ where $\bx \in \mathbb{R}^K$ is the $K$-dimensional vector representing the feature of the sample, $z \in \mathbb{N}^+$ is an integer of the true event time and $b_l \in \mathbb{N}^+$ is a integer of the proposed bid price.

As is illustrated in Figure~\ref{fig:DLF}, the input to our DLF model is the triple $(\bx, z, b_l)$.
For each recurrent unit, we feed the input as $(\bx, b_j)$ where $j \in [1, l]$ is an integer representing the price interval of the current unit.

Note that $\bx$ is a multi-hot encoded feature vector including a series of one-hot encoded features, and we first put it through an embedding layer as
$$ \be = \text{embed}(\bx).$$
After getting the embedding vector $e$, we concatenate the input vector $\be$ and the bid $b_j$ as
$$\bv_j = \text{concat}(\be, b_j).$$
Specifically, we implement Long Short-Term Memory (LSTM)  \cite{hochreiter1997long} as the recurrent unit $f_{\btheta}(\bx, b_j ~|~ \br_{j-1})$ as

\begin{equation}
\begin{aligned}
\textbf{f}_j &= \sigma (\textbf{W}_f \cdot \bv_j + \textbf{b}_f) \\
\textbf{i}_j &= \sigma (\textbf{W}_i \cdot \bv_j + \textbf{b}_i) \\
\textbf{o}_j &= \sigma (\textbf{W}_o \cdot \bv_j + \textbf{b}_o) \\
\br_j = \textbf{f}_j \odot &\br_{j-1} + \textbf{i}_j \odot \text{tanh}(\textbf{W}_s \cdot \bv_j + \textbf{b}_s) \\
\textbf{l}_j &= \textbf{o}_j \odot \text{tanh}(\br_j).
\end{aligned}
\end{equation}
Here $\br_j$ is the hidden state vector of the $j$th recurrent unit.
After we get the output of each unite $\textbf{l}_j$, a fully connected layer with sigmoid activation function predicts the hazard rate as is described in Eq.~(\ref{eq:discrete-hazard-function}) of our main paper as
$$ h_j = \sigma(\textbf{W}_h \cdot \textbf{l}_j + \textbf{b}_h).$$

Then we may calculate the market price probability $p(z)$ w.r.t. market price $z$, the winning rate $W(b)$ and the losing rate $S(b)$ at the proposed bid price $b$ as Eqs.~(\ref{eq:discrete-win-survival-def}) and (\ref{eq:pdf-calc}) of our main paper.
More details can be referred to our published code and the link of code repository with datasets is \href{https://github.com/rk2900/DLF}{https://github.com/rk2900/DLF}.

\section{Experiment Configurations}
\subsection{Hyperparameter}
All the models are trained until convergence and we consider learning rate from $\{1 \times 10^{-4}, 5 \times 10^{-3}, 1 \times 10^{-3}\}$. The value of $\alpha$ is tuned to 0.25. Batch size is fixed on 128 and embedding dimension is 32. All the deep learning models take input features and feed through an embedding layer for the subsequent feedforward calculation. The hyperparameters of each model are tuned and the best performances have been reported.

\subsection{Hardware}
The models are trained under the same hardware settings with an Intel(R) Core(TM) i7-6900K CPU processor, an NVIDIA GeForce GTX 1080Ti GPU processor and 128 GB memory. The training time of each compared model is less than ten hours (as reported from the slowest training model MTLSA) on each dataset.
\end{appendices}

\end{document}